# A Brief Discussion on the Crossovers in Detrended Fluctuation Analysis

Huijie Yang[♣] [1], Fangcui Zhao[1], Xizhen Wu[2], Zhuxia Li[2], Yizhong Zhuo[2]

[1] *Physics Institute, Hebei University of Technology, Tianjin300130, China*
[2] *China Institute of Atomic Energy, P.O. Box 275(18), Beijing102413, China*

**Abstract**

An analytical formula for the contributions of the trend leftovers in DFA method is presented, based upon which the crossovers in DFA are investigated in detail. This general formula can explain the calculated results with DFA method for some examples in literature very well.

[♣] Corresponding author, E-mail: huijieyangn@eyou.com

# I. Introduction

Measurement records for a complex dynamical process form a long time series. And analysis of the time series will reveal the essential dynamical mechanics for the process, which in turn can provide us physical pictures and criteria to construct dynamical models. Many natural sequences have attracted special attentions recently. Typical examples include DNA sequences, weather and climate records, heartbeat and gait series, price evolutions and information streams in Internet, etc. [1-11]. A common feature for all these sequences is that there exist a long-range correlation for the elements, that is to say, the correlation function obeys a power-law and there is not a finite correlation length. When we begin to study these admirably accurate data, almost immediately we encounter three roadblocks [5]. The first is that we must find the non-trivial self-similar properties instead of the trivial self-similar ones. Rescaling the t (time) dimension only and keeping the y (time interval) dimension invariance, we can obtain segments at different time scales. The segments have a self-similar property. But it is a trivial one that we are not interested in. The second is that environments always perturb measurements. And the perturbations can be described with white noise. What is more, statistical noise appears due to the finite number of measurements. The third one is that these data have the property of non-stationary. This means that the statistical properties of a time series are not constant in time; they are not served up neatly as independent, identically distributed random variables. Traditional approaches such as the power-spectrum and correlation analysis are not suited to accurately quantify long-range correlations in this kind of non-stationary signals. Hence, non-stationary is the essential problem to be resolved [12-15].

In literatures [16]，DFA method is designed to solve these problems. DFA is a scaling analysis method providing a simple quantitative parameter一the scaling exponent $\alpha$一to represent the correlation properties of a time series. The advantage of DFA over many methods are that it permits the detection of long-range correlations embedded in seemingly non-stationary time series, and also avoids the spurious detection of apparent long-range correlations that are artifact of non-stationary. In the past few years, more than 100 publications have utilized the DFA as method of correlation analysis, and fruitful results are achieved in many research fields such as cardiac dynamics, bio-informatics, economics, meteorology, geology, etc. [17-19]

The correct interpretation of the scaling results obtained by the DFA method is crucial for understanding the intrinsic dynamics of the systems under study. In fact, just as pointed out in references [17-19], for all systems where the DFA method was applied, there are many issues that remain unexplained. One of the common challenges is that the correlation exponent is not always a constant (independent of scale) and crossovers often exist. A crossover usually can arise from a change in the correlation properties of the signal at different time/space scales, or can often arise from trends in the data. Though DFA-n method can eliminate trends up to $n$-order, the leftovers may generate artifact long-range correlations, which in turn can induce crossovers. In this paper we present a general formula for the leftovers' contribution in DFA method. As an example, a leftover with sinusoidal form is investigated in detail. The results can explain the calculations presented in reference [18,19] exactly.

## II. DFA method and the contribution of leftovers

**1. DFA method**

Normally, to find the non-trivial self-similar properties and reduce the perturbations due to noises, an integrated time series is introduced instead of investigating the initial time series directly. The integrated time series can be constructed as follows,

$$Y_m = \sum_{i=1}^{m} \Delta \tau_i$$

Where $\{Y_m | m = 1,2,3,...\}$ forms the integrated time series, called $Y_m$ "profile", and $\{\Delta \tau_i | i = 1,2,3,...\}$ is the initial time series.

To avoid spurious detection of correlations due to artifact of non-stationary, DFA method is suggested to calculate the time-dependent fluctuation function. The DFA procedure consists of four steps, as illustrated below [19],

***Step.1*** Construct the profile as described above.

***Step.2*** Cut the profile $Y(i)$ into $N_s \equiv [N/s]$ non-overlapping segments of equal length $s$. Since the record length $N$ need not be a multiple of the considered time scale $s$, a short part at the end of the profile will remain in most cases. In order not to disregard this part of record, the same procedure is repeated starting from the other end of the record. Thus, $2N_s$ segments are obtained altogether.

***Step.3*** Calculate the local trend for each segment $\nu$ by a least-squares fit of the data. Then define the detrended time series for segment duration $s$, denoted by $Y_s(i)$, as the difference between the original time series and the fits: $Y_s(i) = Y(i) - P_\gamma(i)$, where $P_\gamma(i)$ is the fitting polynomial in the $\gamma$'th segment. Linear, quadratic, cubic or higher order polynomials can be used in the fitting procedure (called DFA-1, DFA-2, and DFA-n respectively). Since the detrending of the time series is done by subtraction of the fits from the profile, these methods differ in their capability of eliminating trends in the data. In DFA-n trends of order $n$ in the profile and of $n-1$ in the original record are eliminated. Thus, a comparison of the results for different orders of DFA allows estimating the strength of the trends in the time series.

***Step.4*** Calculate the variance for each of the $2N_s$ segments:

$$F_s^2(\nu) = \langle Y_s^2(i) \rangle = \frac{1}{s} \sum_{j=1}^{s} Y_s^2[(\nu-1)s + i],$$

of the detrended time series $Y_s(i)$ by averaging over all data points $i$ in the $\nu$'th segment. Finally, average over all segments and take the square root to obtain the DFA fluctuation function:

$$F(s) = \left[ \frac{1}{2N_s} \sum_{\nu=1}^{2N_s} F_s^2(\nu) \right]^{\frac{1}{2}}.$$

For different detrending orders $n$ we can obtain different fluctuation functions, denoted by $F^{(n)}(s)$.

It can be proved that if the initial data $\{x_i | i = 1,2,...N\}$ are long-range power-law correlated, e.g., $C(s) = \langle x_i x_{i+s} \rangle \propto s^{-\gamma}$, the fluctuation function $F^{(n)}(s)$ increase by a power-law, $F^{(n)}(s) \propto s^{1-\frac{\gamma}{2}} = s^\alpha$, for large $s$ values.

As a brief argument we consider a time series without trends and zero offset. Then the mean-square displacement in each segment $\nu$ can be calculated:

$$\langle Y^2(i) \rangle = \left\langle \sum_{k=1}^{i} x_k^2 \right\rangle + \left\langle \sum_{k \neq j}^{j,k \leq i} x_j x_k \right\rangle = i \langle x^2 \rangle + \sum_{k \neq j}^{j,k \leq i} C(|k-j|) = i \langle x^2 \rangle + 2 \sum_{k=1}^{i-1} (i-k) C(k).$$

And $C(k)$ is the autocorrelation function: $C(s) = \langle x_i x_{i+s} \rangle \propto s^{-\gamma}$.

For large $i$, the second term can be approximated:

$$\sum_{k=1}^{i-1} C(k) \sim \sum_{k=1}^{i} k^{-\gamma} \sim \int_1^i k^{-\gamma} dk \sim i^{1-\gamma}.$$

And

$$\sum_{k=1}^{i-1} kC(k) \sim i^{2-\gamma}.$$

If the data are long-range power-law correlated with $0 < \gamma < 1$, this term will dominate for large $i$, giving:

$$\langle Y^2(i) \rangle \sim i^{2-\gamma}.$$

A similar approximation for $F^{(n)}(s)$ leads to:

$$F^{(n)}(s) \propto s^{1-\gamma/2} = s^{\alpha}.$$

**2. The contribution of the leftovers**

The detrending procedure in DFA method can eliminate the trends effectively, but it cannot separate the dynamical fluctuations from the trends exactly. Actually, the detrended profile contains two parts. One part is the dynamical fluctuations and the other part is the leftovers of the trends, $T(i), i=1,2,..N$.

A good characteristic for the leftovers is that they obey a deterministic function, e.g., they are correlated completely. What is more, because the two parts are independent on each other, the coupling between them should be zero. Denoting the integrated true trend, the eliminated trend in DFA method, the integrated leftovers and the integrated dynamical fluctuations with $T_I(i)$, $T_D(i)$, $\Delta T_I(i) = T_I(i) - T_D(i) = \sum_{n=1}^{i} T(n)$ and $F(i)$ respectively, the calculated result with DFA can be obtained as,

$$\sqrt{\Delta_s^2} = \sqrt{\langle (\Delta T_I(i) + F(i))^2 \rangle} = \sqrt{<\Delta T_I(i)^2> + <F(i)^2> + 2<\Delta T_I(i) \cdot F(i)>}$$

$$= \sqrt{<\Delta T_I(i)^2> + <F(i)^2> + 2<\Delta T_I(i)> \cdot <F(i)>}$$

$$= \sqrt{<\Delta T_I(i)^2> + <F(i)^2>}$$

That is to say, the calculated result with DFA contains two separate contributions. One part is from the dynamical fluctuations and the other part is from the leftovers.

The deterministic trend with length $N$ (e.g., leftovers), denoted with $T(x)|x = 1,2,...N$ here, can be extended periodically to the whole space as,

$$G(x) = G(x + mN) = T(x)|m = 0, \pm 1, \pm 2,...; 1 \le x \le N$$

Accordingly, $G(x)$ can be expressed with the set $\left\{ e^{i \cdot \frac{2n\pi}{N} x} \middle| n = 1,2,3,...N \right\}$, which reads,

$$G(x) = \sum_{n=1}^{N} a_n \exp(i \cdot \tfrac{2n\pi}{N} x)$$

The autocorrelation function for the trend $G(x)$ can be defined as,

$$C(\tau) = c \cdot \int_1^N G(x)^* \cdot G(x+\tau) dx \propto \sum_{n=1}^{N} \sum_{m=1}^{N} a_m^* a_n \int_1^N \exp(-i \cdot \tfrac{2n\pi}{N} x) \exp(\tfrac{2m\pi}{N} x + \tfrac{2m\pi}{N} \tau) dx$$

$$\propto \sum_{n=1}^{N} \sum_{m=1}^{N} a_n^* a_m \delta(n-m) \exp(i \cdot \tfrac{2n\pi}{N} \tau) = \sum_{n=1}^{N} |a_n|^2 \cdot \exp(i \cdot \tfrac{2n\pi}{N} \tau).$$

Therefore,

$$\left\langle (\Delta T(\tau))^2 \right\rangle = \tau \left\langle G(x)^2 \right\rangle + 2 \sum_{k=1}^{\tau-1} (\tau - k) C(k).$$

The first term is,

$$\tau \left\langle G(x)^2 \right\rangle = \tau \cdot \sum_{n=1}^{N} |a_n|^2 = c_1 \cdot \tau$$

The second term is,

$$2 \sum_{k=1}^{\tau-1} (\tau - k) C(k) = 2 \cdot \sum_{k=1}^{\tau-1} (\tau - k) \sum_{n=1}^{N} |a_n|^2 \cdot \exp\left( i \cdot \frac{2n\pi}{N} \cdot k \right)$$

$$= 2 \cdot \sum_{n=1}^{N} |a_n|^2 \sum_{k=1}^{\tau-1} (\tau - k) \cdot \exp\left( i \cdot \frac{2n\pi}{N} \cdot k \right)$$

$$= 2 \cdot \sum_{n=1}^{N} |a_n|^2 \left[ \tau \cdot \sum_{k=1}^{\tau-1} \exp\left( i \cdot \frac{2n\pi}{N} \cdot k \right) - \sum_{k=1}^{\tau-1} k \cdot \exp\left( i \cdot \frac{2n\pi}{N} \cdot k \right) \right]$$

For large $\tau$ we have,

$$\sum_{k=1}^{\tau-1} \tau \cdot \exp\left(i \cdot \frac{2n\pi}{N} \cdot k\right) \sim \tau \cdot \int_{1}^{\tau-1} \exp\left(i \cdot \frac{2n\pi}{N} \cdot k\right) dk = \frac{-N\tau}{2n\pi} i \cdot \left[\exp\left(i \cdot \frac{2n\pi}{N} \cdot (\tau-1)\right) - \exp\left(i \cdot \frac{2n\pi}{N}\right)\right]$$

$$\sum_{k=1}^{\tau-1} k \exp\left(i \cdot \frac{2n\pi}{N} \cdot k\right) \sim \int_{1}^{\tau-1} k \exp\left(i \cdot \frac{2n\pi}{N} \cdot k\right) dk = \frac{d\int_{1}^{\tau-1} \exp\left(i \cdot \frac{2n\pi}{N} \cdot k\right)}{d\left(i \cdot \frac{2n\pi}{N}\right)}$$

$$= \frac{d\left\{\frac{-N}{2n\pi} i \cdot \left[\exp\left(i \cdot \frac{2n\pi}{N} \cdot (\tau-1)\right) - \exp\left(i \cdot \frac{2n\pi}{N}\right)\right]\right\}}{d\left(i \cdot \frac{2n\pi}{N}\right)}$$

$$= \frac{-N}{2n\pi} i \cdot \left[(\tau-1)\exp\left(i \cdot \frac{2n\pi}{N}(\tau-1)\right) - \exp\left(i\frac{2n\pi}{N}\right)\right] - \left(\frac{N}{2n\pi}i\right)^2 \left[\exp\left(i\frac{2n\pi}{N}(\tau-1)\right) - \exp\left(i\frac{2n\pi}{N}\right)\right]$$

Hence, the second term can also be expressed as,

$$2\sum_{k=1}^{\tau-1} (\tau-k)C(k) \propto \sum_{n=1}^{N} |a_n|^2 \left[\frac{-N}{2n\pi} i \cdot \exp\left(i\frac{2n\pi}{N}(\tau-1)\right) + \frac{N(\tau-1)}{2n\pi} i \cdot \exp\left(i \cdot \frac{2n\pi}{N}\right)\right]$$

$$+ \sum_{n=1}^{N} |a_n|^2 \left[\left(\frac{N}{2n\pi}i\right)^2 \cdot \exp\left(i\frac{2n\pi}{N}(\tau-1)\right) - \left(\frac{N}{2n\pi}i\right)^2 \cdot \exp\left(i \cdot \frac{2n\pi}{N}\right)\right]$$

The autocorrelation function should be real number. Therefore we have the following relation,

$$\langle (\Delta T(\tau))^2 \rangle = \tau \langle G(x)^2 \rangle + 2\sum_{k=1}^{\tau-1} (\tau-k)C(k)$$

$$\langle (\Delta T(\tau))^2 \rangle = c_1 \tau + c_t \cdot \sum_{n=1}^{N} |a_n|^2 \left[\frac{N}{2n\pi}\sin\left(\frac{2n\pi}{N}(\tau-1)\right) - \frac{N(\tau-1)}{2n\pi}\sin\left(\frac{2n\pi}{N}\right)\right]$$

$$+ c_t \cdot \sum_{n=1}^{N} |a_n|^2 \left[-\left(\frac{N}{2n\pi}\right)^2 \cos\left(\frac{2n\pi}{N}(\tau-1)\right) + \left(\frac{N}{2n\pi}\right)^2 \cos\left(\frac{2n\pi}{N}\right)\right]$$

The final relation between the statistical quantity $\Delta_\tau^2$ in DFA method and the size of the window $\tau$ can be written as,

$$\Delta_\tau^2 = c_0 \tau + c_f \tau^{2\alpha} + c_t \left\{\sum_{n=1}^{N} |a_n|^2 \left[\frac{N}{2n\pi}\sin\left(\frac{2n\pi}{N}(\tau-1)\right) - \frac{N(\tau-1)}{2n\pi}\sin\left(\frac{2n\pi}{N}\right)\right]\right\}$$

$$+ c_t \left\{\sum_{n=1}^{N} |a_n|^2 \left[-\left(\frac{N}{2n\pi}\right)^2 \cos\left(\frac{2n\pi}{N}(\tau-1)\right) + \left(\frac{N}{2n\pi}\right)^2 \cos\left(\frac{2n\pi}{N}\right)\right]\right\}$$

We can find that the contributions of the leftovers have a complex form.

## III. A Typical Example

As an example, we consider a time series of correlated noise with sinusoidal trend. The leftovers can be expressed as a sinusoidal form. The frequency and the amplitude of the trend can be written as $\frac{2q\pi}{N}$ and $A_s$, e.g., $a_n = \delta(n-q)A_s$. The statistical quantity in DFA method can be reduced to a simple form as,

$$\Delta_\tau^2 = c_0\tau + c_f\tau^{2\alpha} + c_t \cdot A_s^2 \cdot \left[\frac{N}{2q\pi}\sin\left(\frac{2q\pi}{N}(\tau-1)\right) - \frac{N(\tau-1)}{2q\pi}\sin\left(\frac{2q\pi}{N}\right)\right]$$

$$+ c_t \cdot A_s^2 \left[-\left(\frac{N}{2q\pi}\right)^2\cos\left(\frac{2q\pi}{N}(\tau-1)\right) + \left(\frac{N}{2q\pi}\right)^2\cos\left(\frac{2q\pi}{N}\right)\right]$$

The competition between the effects on $\Delta_\tau^2$ of the sinusoidal signal and the correlated noise determines this statistical quantity's scaling behaviors.

When the window size $\tau$ is large enough the effects of triangle terms can be neglected due to their oscillating in a special range. And the effect of the linear term can also be neglected comparing with the term $c_f\tau^{2\alpha}$. Therefore, we can expect that the scaling behavior for $\Delta_\tau^2$ versus $\tau$ should be same with that for pure dynamical fluctuations versus $\tau$, e.g., $\Delta_\tau^2 \sim c_f\tau^{2\alpha}$ in the scale range with large $\tau$.

On the contrary, in the scale range with small $\tau$ (here we consider the condition $\frac{2q\pi}{N}\tau \ll 1$) we can expand the triangle terms into Taylor series and neglect the high order terms. We can find an approximation for the relation between $\Delta_\tau^2$ versus $\tau$ as, $\Delta_\tau^2 = c_0\tau + c_f\tau^{2\alpha} + A_s^2(c_t\tau^2 + ...)$. The relation $F(\tau)^2 \sim c_f\tau^{2\alpha}$ may also be dominant in this condition.

In the intermediate range of the window size $\tau$, the terms in the original relation $\Delta_\tau^2$ versus $\tau$ can be all dominant.

Varying the frequency, e.g., $q$ value and the power-law exponent $\alpha$, we present the results for the relation between $\Delta_\tau^2$ versus $\tau$ in Fig.(1a), Fig.(1b), Fig.(2a) and Fig.(2b). We can find that there are three crossovers in the curves $\Delta_\tau^2$ versus $\tau$, denoted with $n_{1x}, n_{2x}$ and $n_{3x}$ here, respectively. These crossovers separate the whole range of the window size into four parts, denoted with *I, II, III* and *IV* in the figures. The contributions of those triangle terms oscillate with a period, which is same with that of the initial sinusoidal trend. From the figures we can find that the crossover $n_{2x}$ is actually correspondent with the half of the first period. Clearly $n_{2x}$ should be $T/2 = \frac{N}{2q}$ almost exactly.

As for the $n_{1x}$, because it appears in the scale range with small $\tau$, its position should be the intercept of the two terms $c_0\tau + c_f\tau^{2\alpha}$ and $c_t \cdot A_s^2 \cdot \tau^2 + c_t \cdot A_s^2 \cdot \left(\frac{2q\pi}{N}\right)^2 \tau^4 + ...$ in $\Delta_\tau^2 = c_0\tau + c_f\tau^{2\alpha} + c_t \cdot A_s^2 \cdot \tau^2 + ...$. Assuming that the terms $c_f\tau^{2\alpha}$ and $c_t\left(\frac{2q\pi}{N}\right)^2\tau^4$ are dominant respectively in the two terms, we can obtain an estimation of the position of the first crossover as, $n_{1x} = \left(\frac{c_f}{c_t} \cdot \frac{1}{A_s^2} \cdot \left(\frac{N}{2q\pi}\right)^2\right)^{1/4-2\alpha} \sim \left(\frac{T}{A_s}\right)^{1/(2-\alpha)}$. In our results presented here, however, the position of $n_{1x}$ is different with that presented in literature.

The third crossover appears in the scale range with large $\tau$, where two parts may be dominant. One part is the term $c_f\tau^{2\alpha}$ and the other part is the triangle terms. The amplitude of the triangle terms is $A_s^2 \cdot \left(\frac{N}{2q\pi}\right)^2$. Hence $n_{3x} \sim \left(A_s \cdot \frac{N}{2q\pi}\right)^{1/\alpha} \sim (A_s \cdot T)^{1/\alpha}$.

Actually, a DFA-n procedure can eliminate the trends up to (n-1) order, and the leftovers can be a summary of terms with orders $n, n+1,...$. Assuming the term with order $n$ is dominant in the summary, results for high order DFA-n can also be obtained, which is also similar with that presented in reference [18-19].

Comparing Fig.(1a) and Fig.(2a) with Fig.(1b) and Fig.(2b) respectively, we can also find that the effect of a trend with low frequency is much larger than that of a trend with high frequency.

The results above are consistent with the discussions in references [18-19] almost exactly. But from the figures we can find that there are some basically differences comparing with the results in these two references. In the range $\tau < n_{1x}$, the curve for the term obeying power-law and the curve for the total statistical quantity are not coincident with each other, though they can have a similar scaling behavior. While the two curves presented in reference [18] are coincident with each other at all. In the range $n_{1x} < \tau < n_{3x}$, our results possess essentially an oscillating behavior. However, the curves in reference [18] exhibit a flat region for $n_{2x} < \tau < n_{3x}$.

## IV. Conclusions

In summary, based upon random walking theory we present a general formula for the leftovers' contributions in DFA method. A time series of correlated noise with sinusoidal trend is investigated in detail. This formula can explain the calculated results in literatures very well. It may also be helpful for us to understand the DFA results theoretically and estimate the leftovers in DFA analysis procedure. What is more, this formula may provide us useful information to find effective tools for time series analysis.

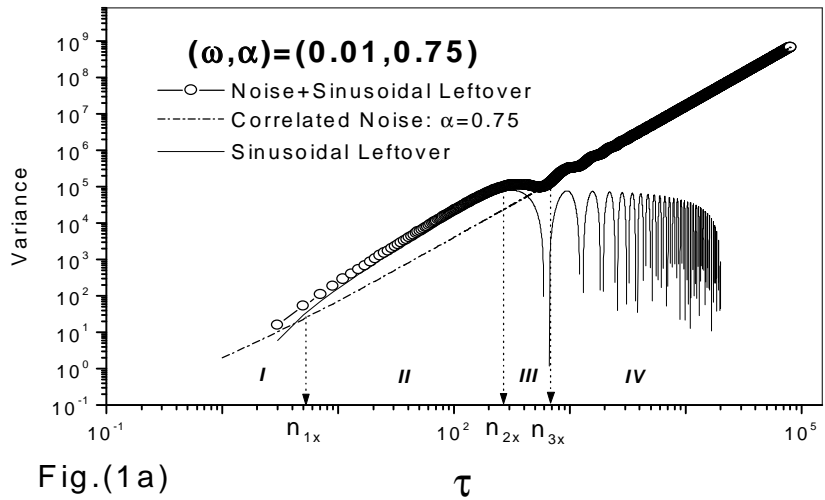

Fig.(1a)

Contributions of the correlated noise with $\alpha = 0.75$ and the trend leftover with sinusoidal form. Here the parameters $(c_0, c_f, c_t)$ are set to be $(1,1,1)$. The amplitude of the leftover is 2. The frequency of the leftover is 0.01.

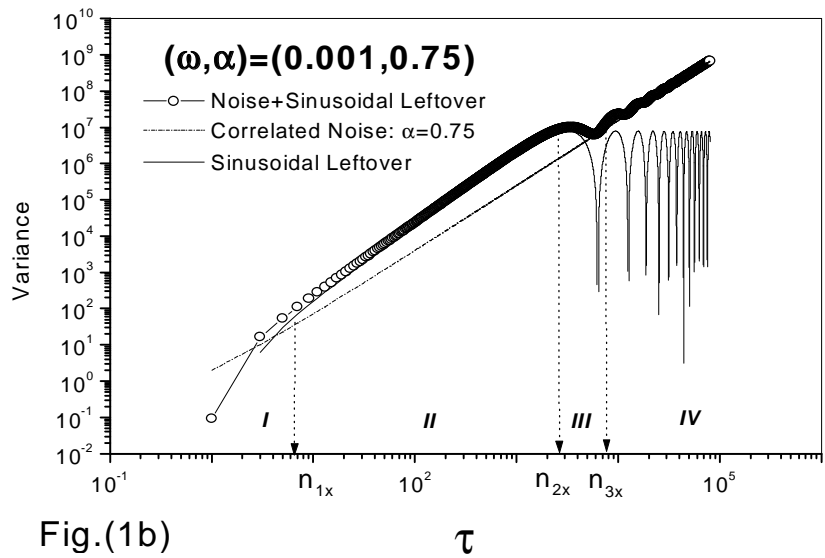

Fig.(1b)

Contributions of the correlated noise with $\alpha = 0.75$ and the trend leftover with sinusoidal form. Here the parameters $(c_0, c_f, c_t)$ are set to be $(1,1,1)$. The amplitude of the leftover is 2. The frequency of the leftover is 0.001.

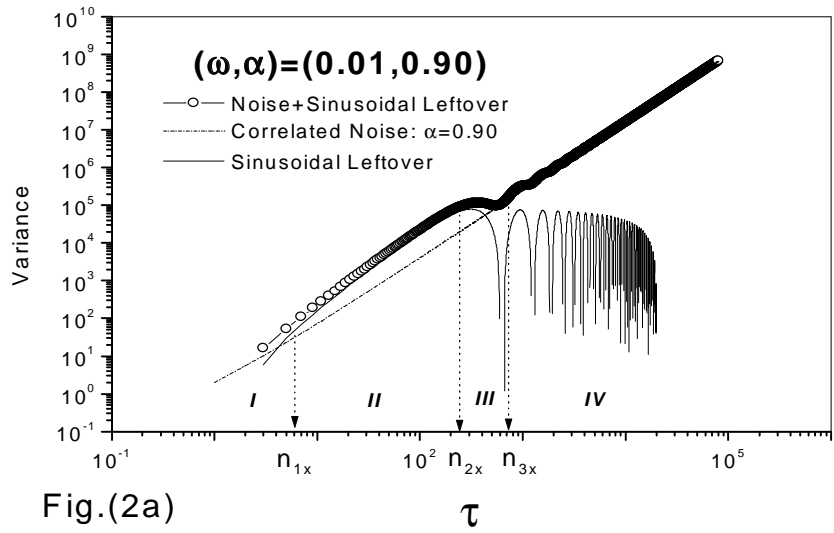

Fig.(2a)

Contributions of the correlated noise with $\alpha = 0.90$ and the trend leftover with sinusoidal form. Here the parameters $(c_0, c_f, c_t)$ are set to be $(1,1,1)$. The amplitude of the leftover is 2. The frequency of the leftover is 0.01.

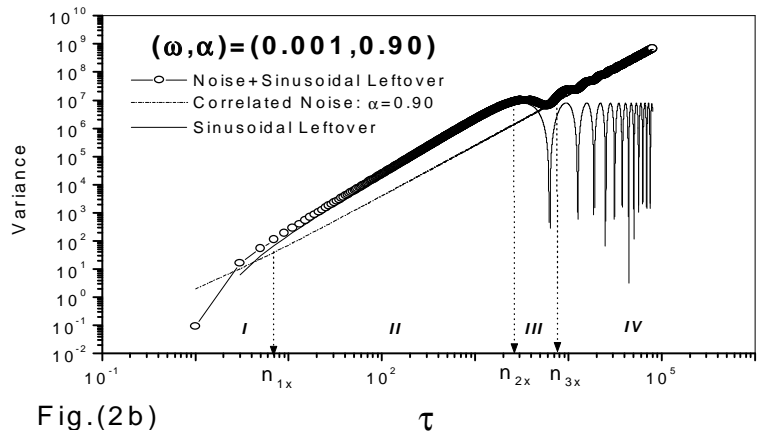

Fig.(2b)

Contributions of the correlated noise with $\alpha = 0.90$ and the trend leftover with sinusoidal form. Here the parameters $(c_0, c_f, c_t)$ are set to be $(1,1,1)$. The amplitude of the leftover is 2. The frequency of the leftover is 0.001.